\documentstyle[prl,twocolumn,aps,epsfig,amsmath]{revtex}

\begin{document}
\draft
\twocolumn[\hsize\textwidth\columnwidth\hsize\csname@twocolumnfalse%
\endcsname

\title{Anomalously large critical regions in power-law 
random matrix ensembles}
\author{E. Cuevas, V. Gasparian$^*$ and M. Ortu\~no}
\address{Departamento de F{\'\i}sica, Universidad de Murcia,
E-30071 Murcia, Spain.}

\date{\today}
\maketitle

\begin{abstract}
We investigate numerically the power-law random matrix ensembles.
Wavefunctions are fractal up to a characteristic length whose logarithm
diverges asymmetrically with different exponents, 1 in the localized
phase and 0.5 in the extended phase. The characteristic length is so
anomalously large that for macroscopic samples there exists a finite
critical region, in which this length is larger than the system size.
The Green's functions decrease with distance as a power law with an
exponent related to the correlation dimension.
\end{abstract}

]
\narrowtext

The power-law random banded matrix (PRBM) model was introduced by 
Mirlin {\it et al.} \cite{Mirlin} to describe a 1D sample with random 
long-range hopping. The model attracted a lot of attention because it
can approximately represent a variety of different physical systems:
an integrable billiard with a Coulomb scattering center \cite{AL97},
two interacting particles in a 1D random potential \cite{PS97},
quantum chaos in a billiard with a non-analytic boundary \cite{CP99},
and the Luttinger liquid at finite temperatures \cite{K99,KS00}.
As the model describes a whole family of critical theories,
it is also interesting for the study of intermediate  spectral
statistics \cite{GJ98,BG99,AJ01}.

The model is represented by $N\times N$ real symmetric matrices whose
entries are randomly drawn from a normal distribution with zero mean
and a variance depending on the distance of the matrix element from
the diagonal
\begin{equation}
\left\langle (H_{ij})^2\right\rangle ={1\over 1+(|i-j|/b)^{2\alpha}} 
\label{hamil}\;.
\end{equation}
From field theoretical considerations \cite{Mirlin,K99,M99,EM00}, 
the PRBM model was shown to 
undergo a sharp transition at $\alpha=1$ from localized states
for $\alpha>1$ to delocalized states for $\alpha<1$. This transition
is supposed to be similar to an Anderson metal-insulator transition,
presenting multifractality of eigenfunctions and non-trivial spectral
compressibility at criticality. The parameter $b$ plays a role analogous
to dimensionality establishing the character of the transition.

From a perturbative treatment of the non-linear $\sigma$-model and
from renormalization group calculations Mirlin considered different
regimes depending on the exponent $\alpha$. These models are only
justified for $b\gg 1$, although the other limiting case can also be
studied following Levitov \cite{levitov}. For $\alpha \ge 3/2$ the
eigenstates are localized, but in contrast to usual
tight-binding models they do not decay exponentially, but with a
power-law with the same exponent $\alpha$ as the hopping matrix
elements \cite{CP99,YO87}.
For $1 < \alpha < 3/2$ the states are still localized, but with a
different typical length. In both cases, the wavefunctions are expected
to have integrable power-law tails and the dimensionality to be equal to
zero. For very large system sizes, the nearest neighbor normalized level 
spacings $s$ should be distributed according to the Poisson law,
$P_{\rm P}(s)=\exp(-s)$.

For $\alpha<1$, the previous theories predict that all states should be 
delocalized, independently of the value of $b$. The inverse participation
ratio (IPR) should then be 
proportional to the system size, as for Gaussian orthogonal ensembles
(GOE), although higher order cumulants of the IPR should not
necessarily behave in the same way as GOE \cite{M99}. For extended states, 
the nearest neighbor level spacing distribution approximately follows 
Wigner-Dyson law \cite{Me91} ,
$P_{\rm WD}(s) = (\pi s/ 2)\exp(-\pi s^2/4)\;.$

At the critical value $\alpha=1$, we expect to have multifractal
eigenstates and critical level statistics, intermediate between
Poisson and Wigner-Dyson. Recent numerical calculations at criticality
and for $b=1$ have shown that the nearest level spacing distribution
differs from the typical one at a metal-insulator transition \cite{VB00}.

Our aim in this paper is to explore in detail the characteristics of the 
PRBM model in the interesting regime around $\alpha\approx 1$ and for the
case $b=1$ where theoretical treatments are not strictly applicable.
We do so by studying systematically the nearest level spacing distribution
function $P(s)$, the behavior of the Green function (GF) as a function of
distance and the fractal properties of the states. We consider the PRBM
Hamiltonian given by Eq.\ (\ref{hamil}) with the exponent $\alpha$ 
ranging between 0.3 and 2. In order to reduce edge effects, we use
periodic boundary conditions, so that our systems are rings of length
$L$. We have checked that the results do not differ significantly from
those obtained for linear chains away from the ends.

We first calculate the eigenvalues of the Hamiltonian matrix by direct 
numerical diagonalization in order to study $P(s)$. We consider a small
energy window $(-0.4,0.4)$ at the center of the band.
For this calculation, the system size varies from $L=200$ to $5000$ and
the number of random realizations is equal to $10^6/L$. We quantitatively
analyze the scaling behavior of $P(s)$ through the following function of
its variance \cite{C99,CL96} 
\begin{equation}
\eta (L, \alpha) = \frac{ {\rm var} (s)- 0.286 }{1-0.286 }\;,\label{eta}
\end{equation}
which describes the relative deviation of ${\rm var} (s)$ from the
Wigner-Dyson limit due to the finiteness of the system. In Eq.\ (\ref{eta})
${\rm var} (s)=\langle s^2\rangle-\langle s \rangle^2$, and 0.286 and 1
are the variances of Wigner-Dyson and Poisson distributions, respectively.
We check that the use of other scaling variables \cite{PV92,SS93}, instead
of $\eta$, do not alter the results.
\vskip -1.0cm
\begin{figure}
\epsfxsize=\hsize
\begin{center}
\leavevmode
\epsfbox{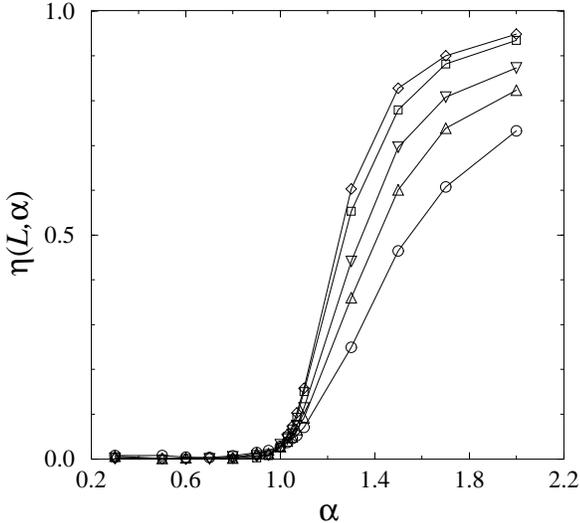}
\end{center}
\caption{Scaling variable $\eta$ as a function of $\alpha$ for different
system sizes: $L=200$ ($\circ$), 500 ($\bigtriangleup$), 1000
($\bigtriangledown$) and 3000
(\protect\raisebox{3pt}{\protect\framebox[7pt]{}}) and 5000 ($\diamond$).
 Up to $\alpha=1$, $\eta$ is basically independent of $L$.}
\label{fig1}
\end{figure}
Figure\ 1 shows the $\alpha$ dependence of $\eta$ for different system 
sizes: $L=200$ ($\circ$), 500 ($\bigtriangleup$), 1000 ($\bigtriangledown$) 
and 3000 (\raisebox{3pt}{\framebox[7pt]{}}) and 5000 ($\diamond$).
The value of $\eta$ at $\alpha=1$ is independent of system size and
very close to the GOE value, instead of being intermediate between
Wigner-Dyson and Poisson like in a standard metal-insulator transition.
This is compatible with the numerical calculations of Ref. \cite{VB00},
where it was found that the spectral statistics at $\alpha= 1$ differs
from that typical of such a transition. For $\alpha> 1$ there is a strong
increase of $\eta$ with both system size and $\alpha$ towards the Poisson
limit. For $\alpha< 1$, $\eta$ is basically independent of system size 
and very similar to the GOE value. The small value of $\eta$ at
criticality makes it difficult to decide whether there is a crossover at
$\alpha=1$, as in standard metal-insulator transitions.
It also indicates that, for $b=1$, the present transition is
characterized by a very large conductance. This characteristic
conductance would decrease with decreasing $b$ \cite{EM00}.

Next, we calculate the real space GF $G(i,j;E)$ as the inverse of the 
Hamiltonian matrix. For each value of $\alpha$ and $L$, we obtain
$G(i,j;0)$ at the center of the band $E=0$ for different disorder
realizations. We first average with respect to the initial position
and then we make an ensemble average of the logarithm of this quantity,
obtaining 
\begin{equation}
F(n)\equiv \langle \ln \sum_i |G(i,i+n;0)|\rangle-\ln L\;.
\end{equation}
In this case, the system size varies from $L=300$ to $4000$ and the 
number of random realizations is equal to $10^6/L$.

In Fig.\ 2 we plot $F(n)-F(0)$ as a function of $n$ for $L=4000$ and
for different values of the exponent $\alpha=0.7$, 0.8, 0.85, 0.9, 0.95 
and 1. We subtract $F(0)$ in order to remove effects associated with 
variations in the average density of states. The central region (tail)
of the results can be fitted very well by a sum of two power-law terms
with the same exponent $\beta$
\begin{equation}
F(n)-F(0)=\ln \left\{{A\over n^\beta}+ 
{A\over (L-n)^\beta}\right\}, \label{a1}
\end{equation}
where $A$ is a constant proportional to the density of states. Each
contribution in Eq.\ (\ref{a1}) corresponds to propagation along one
arm of the ring. 
The exponent $\beta$ is closely related to the matrix exponent $\alpha$
and weakly dependent on system size. The fit of Eq.\ (\ref{a1}) to the
experimental points is so good that it is indistinguishable from the
data on the scale used in Fig.\ 2. Up to the large scales used
states are not truly extended. The results for $\alpha>1$ (not shown)
are similar to those represented in the figure, for $\alpha\le 1$.
\vskip -1.4cm
\begin{figure}
\epsfxsize=\hsize
\begin{center}
\leavevmode
\epsfbox{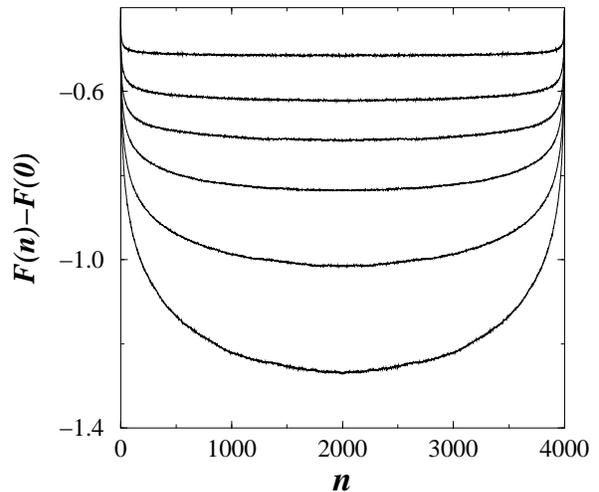}
\end{center}
\caption{$F(n)-F(0)$ as a function of 
$n$ for $L=4000$ and $\alpha=0.7$, 0.8, 0.85, 0.9, 0.95 and 1, from top
to bottom.}
\label{fig2}
\end{figure}
We have tried different extrapolation schemes in order to obtain the
large scale behavior of the exponent $\beta$. We found that a logarithmic
dependence on system size $L$ fits fairly well the results, as can be
seen in Fig.\ 3 where we plot  $\beta$ (solid symbols) versus $L$ for
$\alpha=0.8$, 0.9, 0.95, 1, 1.05 and 1.1 on a semilogarithmic scale. 
The data can be fitted within errors by straight lines.
The value of $\beta$ for the critical case, $\alpha=1$, is size
independent and equal to $0.25$.  For $\alpha<1$, $\beta$ decreases with 
size towards 0, while  for $\alpha >1$ it increases towards $\alpha$.
In the limit of macroscopic sizes, let us say, $L\approx 10^8$, we have
a finite regime where the value of the exponent $\beta$ is different from
zero and so states are not conducting. For infinitely large sizes $\beta$
is zero below the transition, $0.25$ at the transition and $\alpha$ above
the transition.

To clarify the nature of the wavefunctions which give rise to this
anomalous behavior of the GF, we calculate their fractal dimensions
in the region around $\alpha=1$. We obtain the eigenfunctions of the
Hamiltonian matrix by direct numerical diagonalization for system sizes
ranging between $L=500$ and 10000.
For $0.75\alt\alpha \alt 1.25$, we found that the wavefunctions 
show a complex structure with many sharp peaks of different heights
and irregularly spaced which
suggests the possibility of multifractal characteristics. 
We calculate the generalized fractal dimensions $D_q$ by the standard 
box counting procedure through the expression 
\begin{equation}
D_q={1\over q-1}\lim_{\delta\to 0}{\ln\left[\chi_q(l)\right]\over\ln 
\delta}\;, 
\label{as1}
\end{equation}
where $\delta=l/L$ and $\chi_q(l)$ is the $q$-th moment of the 
probability density of the wavefunction in the boxes of size of $l$
\cite{SG91}. For the evaluation of $D_1$ from the previous equation,
one has to expand $\chi_q(l)$ around $q=1$. In practice we compute
Eq.\ (\ref{as1}) at $\delta=0.1$.
\vskip -1.4cm
\begin{figure}
\epsfxsize=\hsize
\begin{center}
\leavevmode
\epsfbox{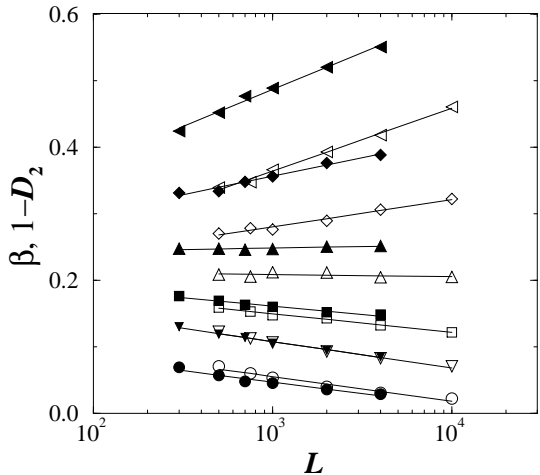}
\end{center}
\caption{Exponent $\beta$ characterizing the decrease with distance 
of the GF (solid symbols) and one minus the fractal dimension $1-D_2$
(empty symbols) as a function of $L$ on a logarithmic scale for
$\alpha=0.8$, 0.9, 0.95, 1, 1.05 and 1.1, from bottom to top. Solid
lines are linear fits to the data.}
\label{fig3}
\end{figure}
We concentrate on the correlation dimension $D_2$, which reflects the
scaling of the density-density correlation function. We found that, as
for the exponent $\beta$, a logarithmic dependence of $D_2$ with $L$
fits fairly well the numerical data. We also found the following
empirical relation between $\beta$ and $D_2$
\begin{equation}
\beta + D_2= \begin{cases}
1 &  \mbox{for $\alpha < 1$}\;, \cr
\alpha & \mbox{for $\alpha \ge 1$}\;. \label{suma}
\end{cases}
\end{equation} To show graphically this relationship, we plot in Fig.\ 3,
together with $\beta$ (solid symbols), the value of $1-D_2$ (empty symbols)
as a function of $L$ for $\alpha=0.8$, 0.9, 0.95, 1, 1.05 and 1.1, from
bottom to top. In agreement with Eq.\ (\ref{suma}), the data for $\beta$
and $1-D_2$ overlap for $\alpha<1$, while they are approximately shifted
by an amount $\alpha-1$ for $\alpha>1$. This shift is consistent with the 
fact that in this regime $D_2$ tends towards zero, while $\beta$ goes to
$\alpha$.

At $\alpha=1$, $D_2$ is practically size 
independent and equal to $0.8$. It very slowly tends to 0 for $\alpha>1$
and to 1 for $\alpha<1$, but even for (finite) macroscopic sizes
remains between 0 and 1 for a finite width interval of the parameter
$\alpha$. The value of
$D_2/d$ at criticality is larger than for the standard Anderson
metal-insulator transition in 3D \cite{SG91}, $1.68/3=0.56$, which
indicates that the characteristic critical adimensional conductance for
the PRBM model with $b=1$ is larger than for the Anderson model. This
is in agreement with the small value found for the variance of the
nearest level spacing at criticality, Fig.\ 1.

As for $D_2$, the other fractal dimensions are size independent at
criticality. We obtained that the information (or entropy) dimension
$D_1$ is equal to 0.88. Other fractal dimensions are $D_{-2}=1.27$,
$D_{-1}=1.14$, $D_{3}=0.71$ and $D_{4}=0.65$. Trivially the similarity
dimension $D_0=d$. In the range of dimensions studied, the deviations
with respect to the weak multifractality limit, where $(D_q/d)-1\propto q$,
are of the order of 10 \%.

Away from the critical point, the wavefunctions are fractal up to a
given length, which depends drastically on $\alpha$ and diverges as
$\alpha$ goes to unity. This length $\xi_{\rm frac}$ is defined as
the system size for which the correlation dimension becomes either 
0 or 1 in a plot like Fig.\ 3. In Fig.\ 4 we show  $\log\xi_{\rm frac}$
as a function of $\alpha$. We see that  $\log\xi_{\rm frac}$ diverges 
at $\alpha=1$ in an
asymmetric way, unlike in standard metal-insulator transitions. 
In the regime $\alpha>1$, a perturbative treatment of the non-linear
$\sigma$-model predicts the existence of a length scale $\xi$
\begin{equation}
\log\xi\sim c{2\alpha-1\over 2\alpha-2} \;,\label{length}
\end{equation} 
which plays the role of the localization length for the PRBM model
\cite{M99,EM00}. Note that $\xi$ is not a length characterizing an 
exponential
decay of the wavefunction, since the GF, for system sizes larger than 
$\xi_{\rm frac}$, decreases with distance as a power law with an exponent
equal to $\alpha$. For this reason it is the logarithm of $\xi_{\rm frac}$
rather than $\xi_{\rm frac}$ itself what diverges in Eq.\ (\ref{length})
with a critical exponent and so plays the role of a localization
length in a standard metal-insulator transition. Eq.\ (\ref{length}) was
obtained for the regime $b\gg 1$ and the constant $c$ was estimated to be
of the order of $\log b$, although one can expect it to be qualitatively
valid for any value of $b$. Our data in the localized regime can be fitted
very well by a curve of the form given by Eq.\ (\ref{length}), as can be
seen in Fig.\ 4, where the right continuous curve corresponds to this
equation with $c=1.67$. This value of $c$ is larger than predicted by
the theory if extended to our case $b=1$.

For $\alpha<1$, the data of Fig.\ 4 cannot be fitted by Eq.\ (\ref{length}),
which corresponds to a critical exponent equal to unity, and there is no
other theoretical prediction available. In this regime the data can be
fitted to a curve of the form
\begin{equation}
\log\xi\sim {c\over |\alpha-1|^\nu} \;,\label{length2}
\end{equation}
with the critical exponent $\nu$ equal to 0.5. The left solid line
in Fig.\ 4 corresponds to Eq.\ (\ref{length2}) with $\nu=0.5$ and
the fitting parameter $c=1.92$. It is the first time to our knowledge
that a metal-insulator transition is approach by two different critical
exponents, which is in contradiction with a single parameter scaling
theory. The possibility of the existence of two characteristic lengths
has been already pointed out in a different one-dimensional model
\cite{DL00}. We note that we are able to obtain two exponents numerically
because the transition point is known exactly ($\alpha=1$), which
constitutes a key advantage of this model.
\vskip -1.4cm
\begin{figure}
\epsfxsize=\hsize
\begin{center}
\leavevmode
\epsfbox{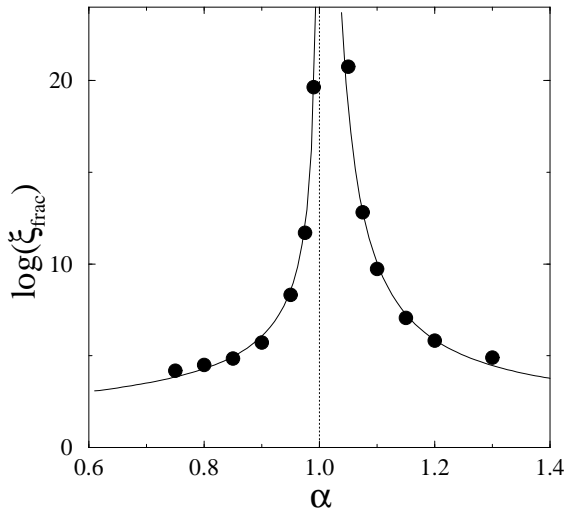}
\end{center}
\caption{Length $\xi_{\rm frac}$ as a function of $\alpha$. Solid 
lines correspond to Eq.\ (\ref{length}) with $c=1.67$, and Eq.\
(\ref{length2}) with $c=1.92$ and $\nu=0.5$.}
\label{fig4}
\end{figure}
To summarize, we have explored the critical region of the PRBM model
in the case $b=1$ inaccessible to field theoretical methods. 
Our results, although compatible with theoretical predictions, show new
light on the specific form of the GF and on the multifractal structure of
the wavefunctions near the critical point.
The wavefunctions are fractal up to a given length which diverges 
asymmetrically with two different critical exponents at $\alpha=1$. 
It is the first time that such an asymmetry has been obtained
in part due to the exact knowledge of the position of the critical 
point.The possibility of a similar asymmetric behavior in other 
metal-insulator transitions should be reconsidered. We note that a
fitting procedure of the raw data without an exact knowledge
of the critical value of the transition parameter is compatible
with equal exponents in the diverging lengths, above and below the
transition.
The fractal behavior is correlated with a power law decay of the GF
with small exponents. The sum of the exponent characterizing the GF
decay and the correlation dimension is approximately equal to unity 
in the localized phase and to $\alpha$ in the extended phase.
All this should have important consequences in the problem of two
interacting particles and in the other problems related to the PRBM
model although their study is clearly outside the scope of this work.

\acknowledgements
We would like to thank the Spanish DGESIC, project numbers
BFM2000-1059 and 1FD97-1358, for financial support. 

\noindent $^*$ Present address: Department of Physics, California
State University at Bakersfield, USA.

\end{document}